\pdfoutput=1
\documentclass[5p, twocolumn, sort&compress]{elsarticle}  
\usepackage{graphicx}
\usepackage{wasysym}
\usepackage{amssymb}
\usepackage{layouts}
\usepackage{abrev-jour-long}
\usepackage[hidelinks]{hyperref}
\usepackage{epsfig,graphics,bm}
\usepackage{color} 
    \definecolor{Blue}{rgb}{0.0,0.0,1.0}
    \definecolor{Red}{rgb}{1.0,0.0,0.0}
    \definecolor{Green}{rgb}{0.0,1.0,0.0}

\journal{Chaos, Solitons, and Fractals}
\begin{document}

\begin{frontmatter}
\title{
Relativistic chaos in the anisotropic harmonic oscillator \tnoteref{t1}
}
\tnotetext[t1]{$\copyright$ 2018. This manuscript version is made available under the CC-BY-NC-ND 4.0 license http://creativecommons.org/licenses/by-nc-nd/4.0/}
%
\author{Ronaldo S. S. Vieira$^{a,b}$}\ead{rssvieira@ime.unicamp.br}
  
\author{Tatiana A. Michtchenko$^{a}$}\ead{tatiana.michtchenko@iag.usp.br}

\address{$^{a}$Universidade de S\~ao Paulo, IAG, 
Rua do Mat\~ao, 1226, Cidade Universit\'aria, 
05508-090 S\~ao Paulo, Brazil}  

\address{$^{b}$Department of Applied Mathematics, University of Campinas, 13083-859 Campinas, SP, Brazil}

\begin{abstract}
The harmonic oscillator is an essential tool, widely used in all branches of Physics in order to understand more realistic systems, from classical to quantum and relativistic regimes. We know that the harmonic oscillator is integrable in Newtonian mechanics, whether forced, damped or multidimensional. On the other hand, it is known that relativistic, one-dimensional driven oscillators present chaotic behavior. However, there is no analogous result in the literature concerning relativistic conservative, two-dimensional oscillators. We consider in this paper different separable potentials for two-dimensional oscillators in the context of special relativistic dynamics. We show, by means of different chaos indicators, that all these systems  present chaotic behavior under specific initial conditions. In particular, the relativistic anisotropic, two-dimensional harmonic oscillator is chaotic.  The non-integrability of the system is shown to come from the momentum coupling in the kinetic part of the Hamiltonian, even if there are no coupling terms in the potential. It follows that chaos must appear in most integrable classical systems once we introduce relativistic corrections to the dynamics. The chaotic nature of the relativistic, anisotropic harmonic oscillator may be detected in the laboratory using Bose condensates in a two-dimensional optical lattice, extending the  recent experiments on the one-dimensional case.
\end{abstract}


\end{frontmatter}
\section{Introduction}

The fundamental properties of a physical theory are best understood by means of simple examples. 
The harmonic oscillator is one of the most simple systems in which we can visualize the basic properties of mechanical theories, from the classical to the quantum regimes. It is present in all branches of Physics, and is a widely used theoretical laboratory to study more complicated phenomena. For instance, the low-energy regime around a stable equilibrium point of a conservative system can be approximated by a set of harmonic oscillators.
Also, more than just a toy model, the harmonic potential is utilized to model systems in diverse fields, such as optomechanics \cite{aspelmeyerEtal2014RvMP}, econophysics \cite{ahnEtal2017EPL}, quantum computation \cite{chiorescuEtal2004Natur}, semiconductors \cite{zawadzki2017JPCM}, among many others, even being the basis for recent discoveries concerning the foundations of quantum theory \cite{boseEtal2018PRL}.

A distinctive feature of the $n-$dimensional harmonic oscillator is that it leads to uncoupled, linear equations of motion in Newtonian mechanics. This is, though, an exceptional case.
It is a well-known result from classical mechanics that a generic perturbation to a separable potential will produce chaos in the corresponding system \cite{tabor1989chaos, lichtenbergLieberman1992}. Pioneering numerical experiments show it by introducing a coupling term in a two-dimensional oscillator \citep{henonheiles1964AJ}. The onset of chaos in conservative systems comes therefore from the non-linearities and coupling terms in the potential. On the other hand, if the potential is separable, then the system is integrable and the general solution of the equations of motion can be formally found via the Hamilton-Jacobi equation \cite{tabor1989chaos}.


Chaos in Newtonian mechanics is directly related to the properties of the potential.
On the other hand, in special relativistic dynamics there is an intrinsic coupling which does not depend on the potential term; it appears in the kinetic part of the Hamiltonian.
Indeed, in units where $c=1$, the Hamiltonian for a test particle reads
\begin{equation}\label{eq:H}
H = \sqrt{p^2 + m^2} + V(q_i)
\end{equation}
where $m$ is the mass of the particle, $q_i$ are its Cartesian coordinates, $p_i$ are the corresponding canonical momenta and $p^2 = \sum p_i^2$ is the particle's total momentum.  Here, $V(q_i)$ is the potential term coming from Newtonian mechanics. The equations of motion are
\begin{equation}\label{eq:dpidt}
\frac{dp_i}{dt} = -\frac{\partial V}{\partial q_i},
\end{equation}
\begin{equation}\label{eq:dqidt}
\frac{dq_i}{dt} = \frac{p_i}{\sqrt{m^2 + p^2}}
\end{equation} 
(we adopt $m = 1$ along the paper).
These equations are valid in the inertial reference frame in which $V$ is constant in time. In the case of oscillators, as we will see below, it means that the origin of the spatial coordinates is an equilibrium point of the system. The presence of the squared norm $p^2$ in the equations of motion (\ref{eq:dqidt}), i.e. the momentum coupling in the kinetic part of the Hamiltonian (\ref{eq:H}), gives rise to this relativistic, nonlinear coupling term \cite{kimLee1995PRE}.



We investigate in this paper the effects of relativistic momentum coupling in simple mechanical systems. 
In the relativistic domain, the so-called (one-dimensional) ``relativistic harmonic oscillator'', with $V(q)=k\,q^2/2$, was studied by several authors, both in classical and quantum regimes \cite{kimLee1995PRE, lee1995LNP, harvey1972PRD, moreauEtal1994AmJPh, homorodean2004EPhL, liEtal2005JMP, santosEtal2006BJP, poszwa2014AcPPA, petrov2016EJPh}. Although this oscillator is not ``harmonic'' in the strict sense (the period of oscillation depends on energy \cite{harvey1972PRD}), we will keep this terminology, which is the most used in the literature.
Regarding driven oscillators, although the one-dimensional, forced harmonic oscillator is  integrable in Newtonian mechanics, it is known that its relativistic version presents chaotic behavior due to the the nonlinear form of Eq.~(\ref{eq:dqidt}) \cite{kimLee1995PRE, lee1995LNP}.
Other one-dimensional relativistic oscillators also present the onset of chaos when driven by time-dependent forces \cite{schieveHorwitz1991PhLA, kimLee1995PREresonances}.
Since these are one-dimensional systems, an external, time-dependent force is necessary to generate chaotic dynamics. Therefore it is of great interest to verify if chaotic behavior can also appear in conservative, $n$-dimensional Hamiltonian systems (without external forces) due to the momentum coupling, for systems which are integrable according to Newtonian mechanics. We will restrict ourselves to the simplest class of systems: two-dimensional oscillators with uncoupled potentials, and in particular quadratic (``harmonic'') potentials. This is the main goal of our work.

The paper is organized as follows. In Section~\ref{sec:nonlinear} we show the onset of chaos in relativistic two-dimensional, non-linear oscillators, due to the momentum coupling. Section~\ref{sec:harmonic} shows that, even for a two-dimensional harmonic oscillator, chaos appears in the relativistic regime whenever the system is anisotropic. Section~\ref{sec:discussion} summarizes our results and presents our conclusions.



\section{Chaos in relativistic non-linear oscillators}
\label{sec:nonlinear}

We show below that, in Hamiltonian systems with two degrees of freedom, potentials which are integrable in Newtonian mechanics give rise to chaotic systems in special relativity and that, in this case, the non-integrability of the corresponding relativistic system comes solely from the momentum coupling in the equations of motion. 
We give examples from the most simple class of conservative two-dimensional systems: two-dimensional oscillators, given by separable potentials of the form
\begin{equation}\label{eq:separable}
V(x,y) = U_x(x) + U_y(y),
\end{equation} 
with $(q_i) = (x,y)$, where the functions $U_x$ and $U_y$ define the type of oscillator considered. In Newtonian dynamics, motion in separable potentials is always integrable. If chaos is detected in these systems in the relativistic regime, it must be due to the momentum coupling.
The coupling term involving the canonical momentum variables would generate, as in canonical perturbation theory, a chain of resonance islands for small $p^2$, according to the Poincar\'e-Birkhoff theorem. As $p^2$ increases, the separatrices of these resonances would break into homoclinic tangles, destroying the corresponding KAM tori and leading to the onset of chaos \cite{lichtenbergLieberman1992}.
Moreover, the onset of chaos in these systems will indicate that the vast majority of relativistic systems (not only oscillators) will present chaotic behavior due to purely relativistic effects, even if their Newtonian limit is integrable.



\begin{figure}[ht]
\begin{center}
\epsfig{figure=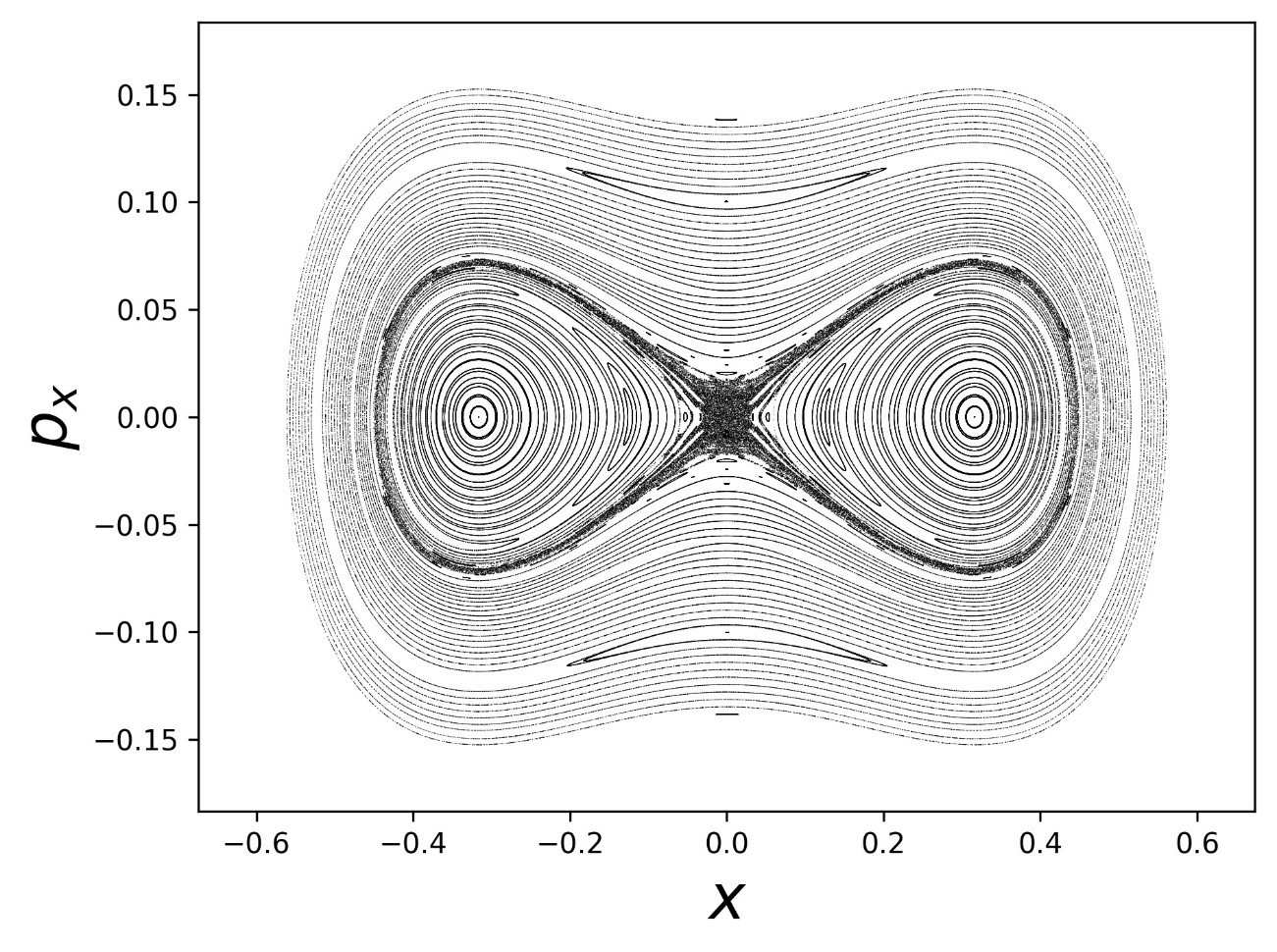,width=0.99\columnwidth ,angle=0}
\caption{Poincar\'e section in the ($x, p_x$) plane with $y=0$ and $p_y>0$, for $E = 1.01$, corresponding to a characteristic speed $v\approx 0.14$. The system corresponds to the Duffing-like oscillator (\ref{eq:AnarmOsc}), with $k_x = k_y = -0.1$ and $w_x = w_y = 1$. The central region is chaotic, corresponding to the hyperbolic nature of the  central fixed point.
}
\label{fig:poincareOMHax-01ay-01bx1by1E1-01}
\end{center}
\end{figure}

Let us first consider a potential with quartic terms
\begin{equation}\label{eq:AnarmOsc}
V_{Q}(x,y) = \frac{1}{2}k_x\,x^2 + \frac{1}{2}k_y\,y^2 + \frac{1}{4}w_x\,x^4 + \frac{1}{4}w_y\,y^4,
\end{equation}
with $k_x,\, k_y,\, w_x,\, w_y$ constants. For $k_x,\, k_y <0$ we have a Duffing-like oscillator \cite{kimLee1995PREresonances}, and, for $k_x,\, k_y >0$, we have a quartic (anharmonic) oscillator.
Regarding the Duffing-like oscillator, we have that, due to the presence of a saddle-like point at the phase-space origin, any additional coupling term in the potential will give rise to chaotic dynamics in the neighborhood of this point. We see in the Poincar\'e section of Fig.\,\ref{fig:poincareOMHax-01ay-01bx1by1E1-01} that the same is true if we add, instead of a coupling term in the potential, the momentum coupling due to relativistic dynamics, Eq.~(\ref{eq:dqidt}). Moreover, the chaotic region appears in this case for relatively low energies, corresponding to characteristic speeds $v\approx 0.1$.

\begin{figure}[ht]
\begin{center}
\epsfig{figure=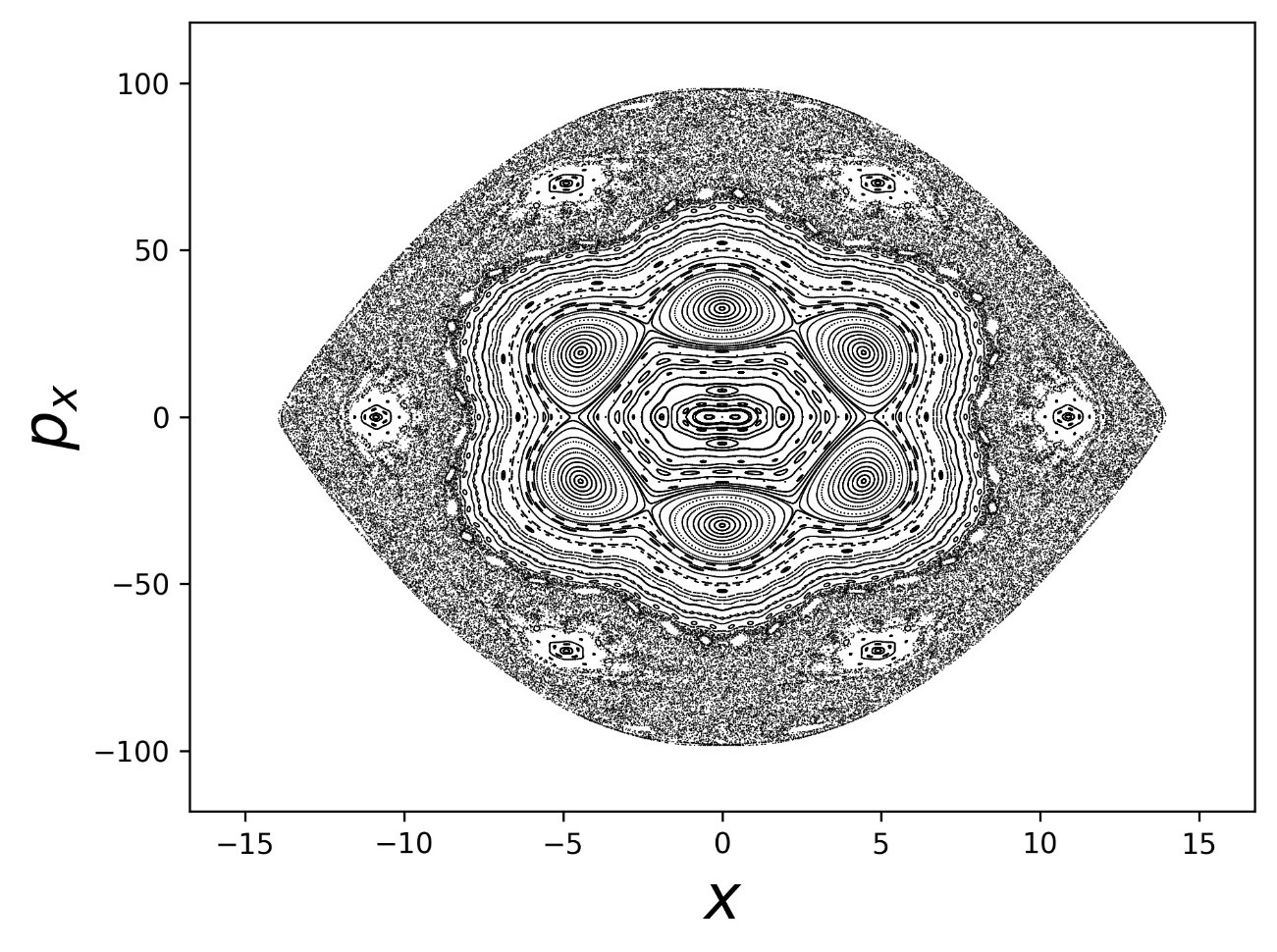,width=0.99\columnwidth ,angle=0}
\caption{Poincar\'e section in the ($x, p_x$) plane with $y=0$ and $p_y>0$, for $E = 100$, corresponding to a characteristic speed $v\approx 0.99995$. The system corresponds to quartic oscillator, Eq.\,(\ref{eq:AnarmOsc}), with $k_x = k_y = 1$, $w_x = 0$, and $w_y = 0.05$. We see a wide region of chaos denoted by the erratic scattering of points filling the region far from (0,0).
}
\label{fig:poincareOQkx1ky1wx0wy00-5E100}
\end{center}
\end{figure}

We also seek chaos in the quartic anharmonic oscillator, Eq.\,(\ref{eq:AnarmOsc}) with $k_x,\, k_y>0$. Since the corresponding Hamiltonian has a stable equilibrium point at the phase-space origin, it is reasonable that chaos does not appear around this point, but instead near the resonance separatrices generated by the relativistic momentum coupling. 
We show a Poincar\'e section for the quartic oscillator in Fig.\,\ref{fig:poincareOQkx1ky1wx0wy00-5E100}, with a wide chaotic region far from $(x, p_x) = (0,0)$. Chaos is found in this system for higher energies than for the Duffing-like potential, corresponding to higher (relativistic) characteristic speeds of the particle. 
The anharmonicity is introduced only in the $y-$ direction, and with a small value $w_y>0$. Note that $w_y$ is also the only parameter which introduces anisotropy in the system, since we take $k_x = k_y = 1$.
Therefore, even a small quartic term in the potential of Eq.\,(\ref{eq:AnarmOsc}) gives rise to wide regions of chaos. 

Up to now, we considered quartic potentials, which give rise to nonlinear equations of motion in Newtonian dynamics. For these relativistic systems, chaos appears exclusively due to the momentum coupling, Eq.\,(\ref{eq:dqidt}), as explained in the beginning of this Section.
In this way, a natural question arises: Could this momentum coupling produce chaos in even simpler relativistic systems?

\section{Chaos in the anisotropic harmonic oscillator}
\label{sec:harmonic}

In order to answer the above question, we consider the simplest two-dimensional potential, the harmonic oscillator
\begin{equation}\label{eq:HO}
V_{HO}(x,y) = \frac{1}{2}\,k_x x^2 + \frac{1}{2}k_y\,y^2, 
\end{equation}
which gives rise to linear, uncoupled equations of motion in the Newtonian formalism (and is therefore integrable). The relativistic equations of motion for $p_i$ (\ref{eq:dpidt}) are also linear, and therefore the only nonlinear term which appears in the equations is the momentum coupling in the equations for $q_i$ (\ref{eq:dqidt}).
Since the isotropic case is integrable, we consider here the relativistic, anisotropic harmonic oscillator, Eq.~(\ref{eq:HO}) with $k_y \neq k_x$.
We find in this section that the system is, in general, non-integrable, in such a way that a generic  $k_y \neq k_x$ will give rise to chaotic dynamics. Therefore, in spite of the linearity of the potential $V_{HO}$, the momentum coupling in the relativistic equations of motion introduces chaotic behavior in the system. We detail below our results.

\begin{figure}[h]
\begin{center}
\epsfig{figure=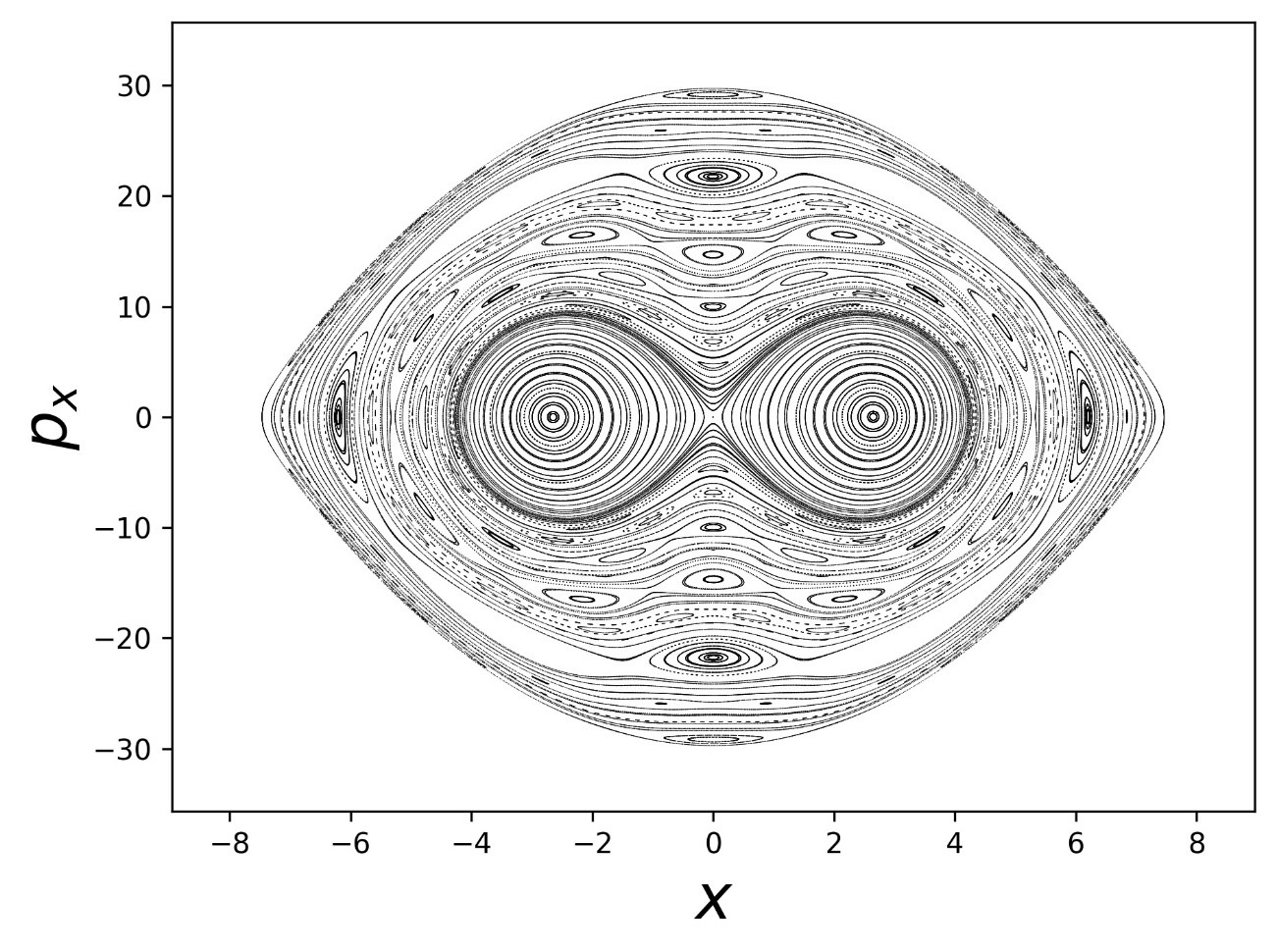, width=0.99\columnwidth ,angle=0}
\caption{Poincar\'e section in the ($x, p_x$) plane with $y=0$ and $p_y>0$, for $E = 30$, corresponding to a characteristic speed $v\approx 0.9994$. The system corresponds to an anisotropic harmonic oscillator (\ref{eq:HO}) with $k_x = 1$, $k_y = 1.5$. The central fixed point $(x, p_x) = (0,0)$ is hyperbolic and corresponds to an unstable periodic orbit. However, no chaos is seen on this scale.
}
\label{fig:poincareOHkx1ky1-5E30}
\end{center}
\end{figure}

\begin{figure}[h]
\begin{center}
\epsfig{figure=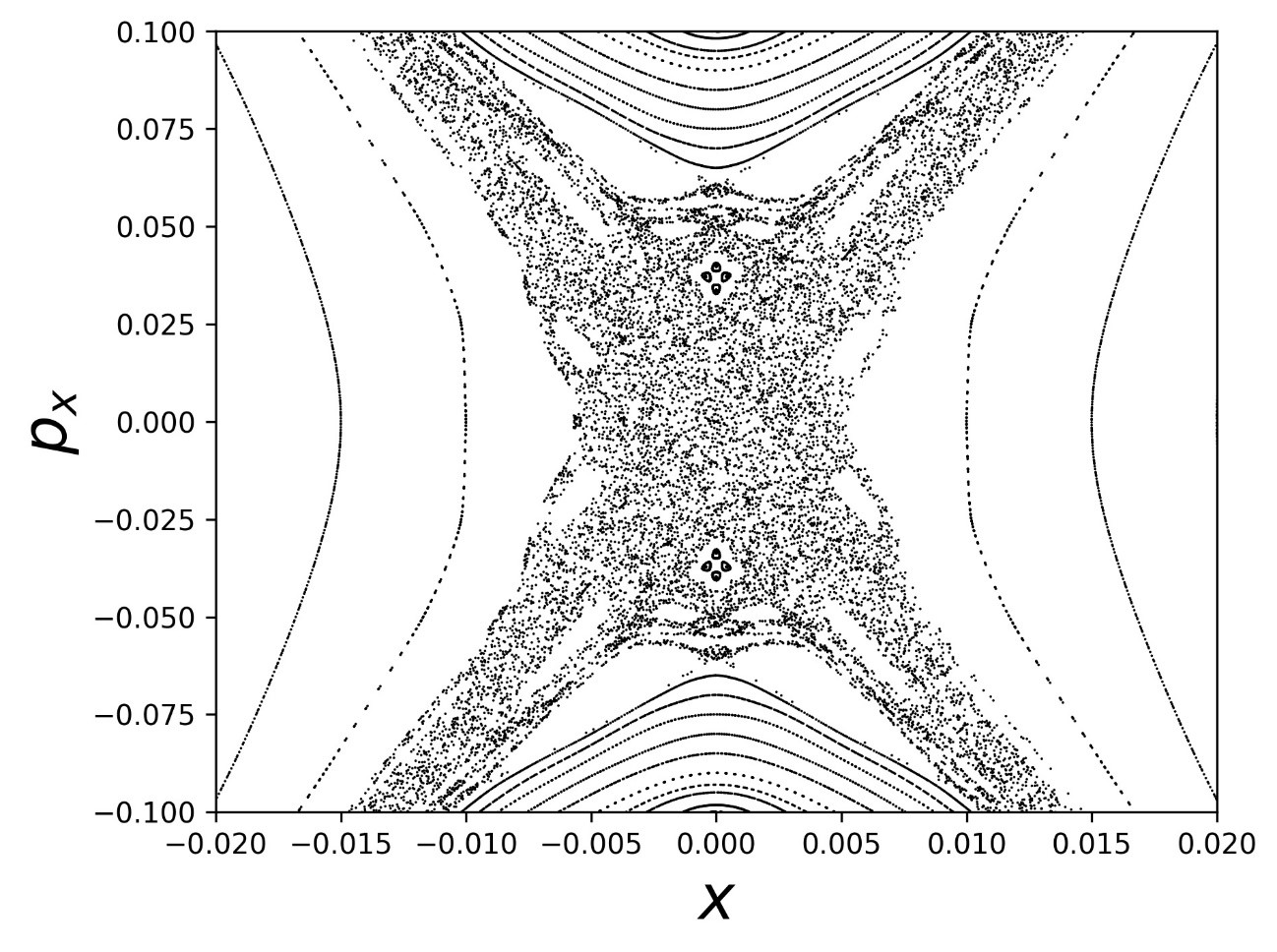,  width=0.99\columnwidth ,angle=0}
\caption{Zoom of the Poincar\'e section of Fig.~\ref{fig:poincareOHkx1ky1-5E30} around $(x, p_x) = (0,0)$; we see on this scale the chaotic behavior due to the motion around the separatrix.
}
\label{fig:poincareOHkx1ky1-5E30x0-2px0-5Zoom}
\end{center}
\end{figure}

Chaos appears around the fixed point $(x,p_x)=(0,0)$ of Poincar\'e sections when this point becomes hyperbolic, corresponding to an unstable periodic orbit in the $y-p_y$ plane (the origin of this instability will be discussed later). We present in Fig.\,\ref{fig:poincareOHkx1ky1-5E30} a Poincar\'e section for the harmonic oscillator, Eq.~(\ref{eq:HO}), with $k_x = 1$, $k_y = 1.5$, in which we can clearly see the hyperbolic nature of the central fixed point, in addition to a whole chain of resonances. This rich resonant structure does not appear in Newtonian mechanics for the system considered, being thus a purely relativistic effect. Although chaos is not visible on this scale, the zoom-in of the central region presented in Fig.\,\ref{fig:poincareOHkx1ky1-5E30x0-2px0-5Zoom} shows a chaotic domain surrounding the hyperbolic central fixed point. 
The erratic scattering of points in this region of the Poincar\'e section is due to the homoclinic tangle between the stable and unstable manifolds associated with the hyperbolic fixed point, which is an imprint of chaotic motion \cite{tabor1989chaos, lichtenbergLieberman1992}. This phenomenon is absent in the Newtonian regime (which is integrable), being thus associated with the relativistic coupling terms. 
This fact leads us to a deeper investigation of the dynamics around this point (which corresponds to a periodic orbit in the $y-p_y$ plane).

\begin{figure}[ht]
\begin{center}
\epsfig{figure=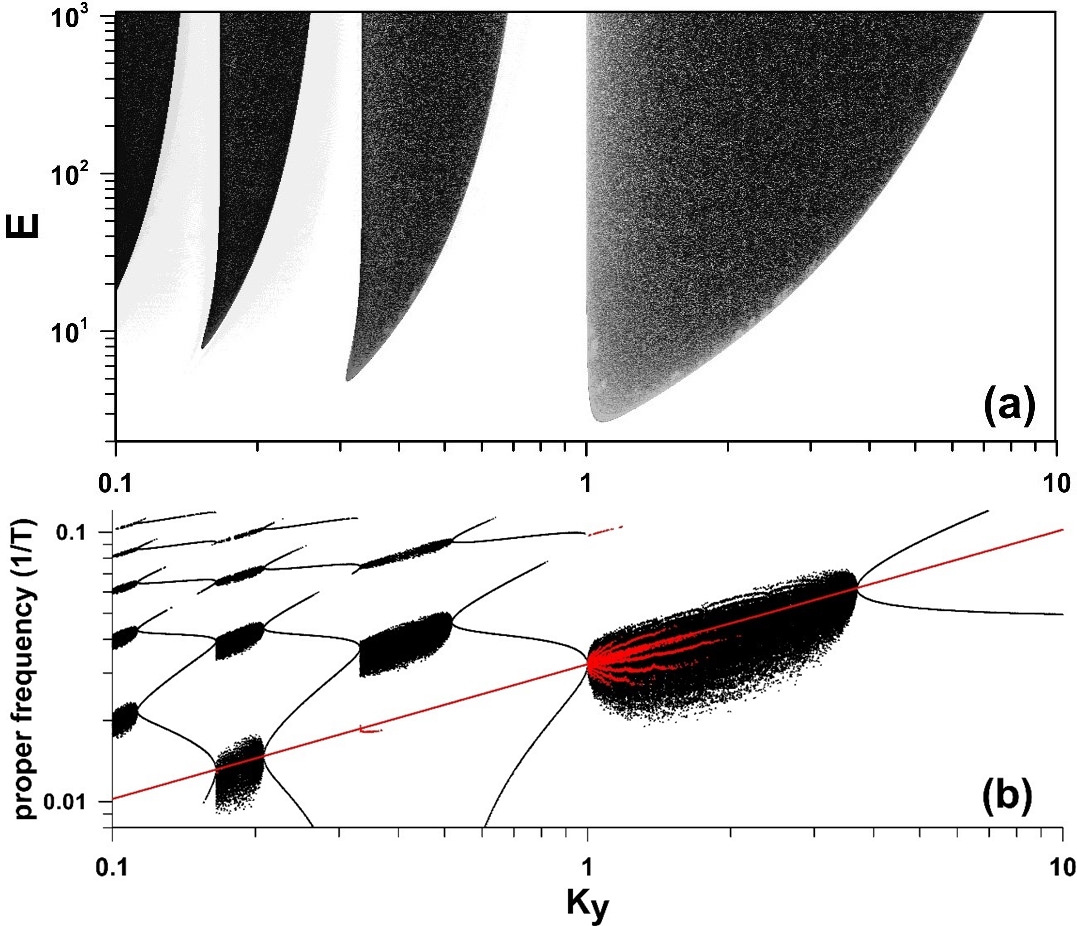, width=0.99\columnwidth ,angle=0}
\caption{(a) Dynamical map for the time series $x(t)$ of the  orbit in the anisotropic harmonic potential, Eq.\,(\ref{eq:HO}), with initial conditions $x = 10^{-20}$, $y = 0$, $p_x = 0$, and $p_y>0$ determined by $E$. The map is calculated on the $k_y-E$ plane (we fix $k_x = 1$). White and light-gray regions correspond to regular orbits, while chaotic orbits are shown by dark regions. Each value of $k_y$ and $E$ also determines the corresponding periodic orbit on the $y-p_y$ plane, which has initial condition $x = 0$. 
The chaotic pattern for $k_y>1$ repeats itself for regions with $k_y<1$. 
(b) Dynamical power spectrum of the orbits with $E = 30$ for the same time series, as a function of $k_y$. The proper frequencies of the system appear as the $x-$mode (black) and the $y-$mode (red), as well as their harmonics and linear combinations. The chaotic regions correspond to resonances between the $x-$ and $y-$ modes: 1/1, 2/1, 3/1 and 4/1, from right to left.
}
\label{fig:DynMap_ky_E}
\end{center}
\end{figure}

The nature of chaos in this system is studied by means of dynamical maps \cite{michtchenkoEtal2002Icar, michtchenkoVieiraBarrosLepine2017AA} in the very close vicinity of the periodic orbit which corresponds to the center of the previous Poincar\'e sections. We present in  Fig.\,\ref{fig:DynMap_ky_E}\,(a) a dynamical map on the $k_y-E$ plane (with $k_x=1$ fixed, without loss of generality), for the orbit given by the initial conditions $x=10^{-20},\, p_x=0,\, y=0$, and $p_y>0$ obtained from the energy $E$. Each point on the plane corresponds to an orbit with the specified initial conditions, for the given parameters of the system. 
The gray scale represents the number of significant peaks on the orbit's power spectrum, connected with the regularity or chaoticity of the orbit \cite{powellPercival1979JPhA}. White regions correspond to regular motion, while chaotic orbits are presented by the black regions.
As $k_y$ increases from 1 to 10, the chaotic region appears in Fig.\,\ref{fig:DynMap_ky_E}\,(a) for higher energies. Moreover, we note that, if we consider instead the same $k_y-E$ plane but with initial condition $x=0$ for the orbits, we have the mentioned periodic orbits on the $y-p_y$ plane (corresponding to the central fixed point of the previous Poincar\'e sections on the $x-p_x$ plane). These periodic orbits are unstable for the dark regions on the map (corresponding to the hyperbolic nature of the central fixed point in the Poincar\'e sections) and stable for the white and light gray regions. We also see that similar structures appear for $k_y<1$. We will show below that these chaotic regions are associated with resonances between the motions on the $x-$ and $y-$ directions. 
In this way, for a wide range of values of $k_y$, a chaotic region exists around these (unstable) periodic orbits for a considerable energy range (see Fig.\,\ref{fig:DynMap_ky_E}\,(a)). 
The system is therefore non-integrable for this range of parameters. 

We interpret the above results, obtained via the dynamical map, by constructing the dynamical power spectrum \cite{michtchenkoEtal2002Icar, michtchenkoVieiraBarrosLepine2017AA} of Fig.\,\ref{fig:DynMap_ky_E}\,(b). It shows the proper frequencies of the orbits (as well as their harmonics and linear combinations), present in their power spectra, as a function of $k_y$ for the same orbits of Fig.\,\ref{fig:DynMap_ky_E}\,(a), but with fixed energy $E = 30$. We see two modes of oscillation. The $y-$mode (red) shows regular behavior and its frequency grows approximately with $k_y^{1/2}$, as expected. The $x-$mode (black) has its fundamental frequency roughly constant (approximately at a value of 0.04), since $k_x = 1$ is fixed. The corresponding time series shows chaotic behavior at resonances between the $x-$ and $y-$ modes, identified by the erratic scattering of frequencies, mainly at the 1/1 resonance (for $k_y>1$). The chaotic regions due to the 2/1, 3/1 and 4/1 resonances also appear for $k_y<1$, from right to left. We may compare the scattering in frequencies for $k_y = 1.5$ with the central chaotic region of Fig.\,\ref{fig:poincareOHkx1ky1-5E30x0-2px0-5Zoom}, both being chaos indicators for the region surrounding the corresponding periodic orbit.

We also calculated the corresponding largest Lyapunov exponent (LLE) \cite{tabor1989chaos, benettinEtal1980Mecc2} for the periodic orbits above. The results obtained have the same qualitative behavior of the dynamical map: positive LLEs were obtained in the dark regions of the $k_y - E$ plane (Fig.\,\ref{fig:DynMap_ky_E}\,(a)), corresponding to unstable periodic orbits surrounded by a chaotic region, and zero values of the LLE were obtained in the regular regions. Also, as expected, we see from Fig.\,\ref{fig:DynMap_ky_E}~top that the Newtonian limit $E\approx 1$ is integrable (The minimum energy for which chaos appears around the periodic orbit is $E\approx 1.4$, corresponding to a characteristic speed $v\approx 0.7$.)

We remark that the above analysis takes into account only the stability of one periodic orbit (for each value of $E$ and $k_y$); there are, in general, many other chaotic domains in phase space. Separatrices generated by the momentum coupling will also break into chaotic regions, in such a way that it is likely that the system is non-integrable for a generic $k_y\neq k_x$, i.e., whenever the oscillator is anisotropic.

\section{Discussion}
\label{sec:discussion}

We considered here the special relativistic dynamics of a particle in a two-dimensional potential well. The potentials are taken as separable and represent two-dimensional oscillators. Although very simple in their nature, all systems considered here were shown to be non-integrable. 
It is remarkable that even such a simple system as the (anisotropic) harmonic oscillator, which would result in linear, uncoupled differential equations in Newtonian mechanics, is also prone to chaos when relativistic corrections to the dynamics are taken into account. Therefore, integrability in special relativistic dynamics is not a property which depends only on the potential; the momentum coupling in Eq.~(\ref{eq:dqidt}) plays a crucial role in the chaotic nature of relativistic motion. The non-integrability of the anisotropic harmonic oscillator comes indeed from this momentum coupling, since the terms in the equations of motion derived from the harmonic potential are all linear. Chaos then appears in relativistic systems regardless of the nonlinearities or coupling terms in the potential.

These results lead us to the following conclusion: A generic, $n$-dimensional integrable Hamiltonian system in Newtonian mechanics is prone to chaos when the characteristic speed of the particle becomes comparable with the speed of light, that is, whenever special relativistic effects are taken into account in the dynamics. Since the potentials analysed in this paper are very simple in their nature, more complicated systems will generally introduce additional resonance islands, whose separatrices would also break into chaotic regions due to the relativistic momentum coupling in the equations of motion. 
More precisely, the expansion of the Hamiltonian, Eq.\,(\ref{eq:H}), in powers of $p^2$ beyond the first-order term $p^2/(2m)$, shows that the system can be written as 
\begin{equation}
H \approx H_N + \delta H,
\end{equation}
where  $H_N = p^2/(2m)+V$ and $\delta H$ contains the higher-order terms in the momentum expansion. If $H_N$ is integrable, $\delta H$ may be considered as a perturbation to the system (although it cannot be considered ``small'' for high characteristic speeds of the particle). Although the perturbation is not generic (it is given by the power series expansion of the kinetic part of the Hamiltonian), it seems unlikely that the system would remain integrable order by order.  

In canonical perturbation theory, the perturbation term in $H$ does not have to be associated only with the potential, but instead it is a function of the phase-space variables \cite{lichtenbergLieberman1992}. It follows that all the results of perturbation theory which are valid for perturbed potentials in classical mechanics will also be valid for the relativistic ``perturbation'' due to the momentum coupling. 
This perturbation leads to small divisors in canonical perturbation theory when the frequencies associated with the fundamental modes of motion in the system are commensurable, which will split the corresponding rational KAM tori into resonance islands. The size of these islands  will grow with increasing perturbation (which in our case is obtained by increasing the system's energy), breaking the KAM tori associated with beating frequencies. The motion around the separatrices of these resonance islands then leads to the onset of chaos, as explained in the text.

These results may have impact on other areas of Physics, such as the study of the effect of classical chaos in quantum Klein-Gordon fields subject to anisotropic quadratic potentials, which may be compared with the isotropic case \cite{poszwa2014AcPPA}.
The momentum coupling also appears in general relativistic test-particle motion if we consider the isoenergetically reduced Hamiltonian (for a given coordinate system and parametrized by coordinate time $x^0=t$)  \cite{chiconeMashoon2002CQGra, vieiraRamoscaroSaa2016PRD}  
\begin{equation}\label{eq:Hred-GR}
   H_{GR} = \frac{g^{0i}p_i}{g^{00}} + \bigg[\frac{1 + \tilde{g}^{ij}p_i p_j}{(-g^{00})}\bigg]^{1/2},
  \end{equation}
in geometrized units and with $m=1$, where $\tilde{g}^{ij}$ is the inverse of the spatial metric ${g}_{ij}$ ($i,j=1,2,3$).	
This Hamiltonian gives rise to the geodesic equations and may be regarded as the general relativistic extension of the special relativistic Hamiltonian (\ref{eq:H}), where instead of the potential $V$ we have the additional coupling terms given by the metric $g_{\mu\nu}$ ($\mu,\nu=0,1,2,3$), which represents the gravitational interaction. This coupling between the metric components is intrinsic to general relativity (in special relativity we have the flat metric $g_{\mu\nu}=\eta_{\mu\nu}$ and therefore the Hamiltonian (\ref{eq:Hred-GR}) reduces to that of a special relativistic free particle (\ref{eq:H}), i.e. in the absence of the external potential). 
Chaos in general relativity occurs in many different astrophysical scenarios, both in exact solutions \cite{saaVenegeroles1999PhLA, wuZhang2006ApJ, semerakSukova2010MNRAS} and in Post-Newtonian expansions \cite{wuXie2007PRD, wangWu2011CQG, huangNiWu2014EPJC, huangWuMa2016EPJC}. 
In this way, we conclude that the momentum coupling presented here is more fundamental to the chaoticity of relativistic systems than the metric coupling, since the latter appears only in general relativity (where the momentum and metric couplings cannot be considered separately) but the coupling between momentum coordinates is already present in special relativistic test-particle motion.

Moreover, the chaotic nature of the classical relativistic, two-dimensional anisotropic harmonic oscillator may be detected experimentally in the near future using Bose condensates in a two-dimensional optical lattice (whose excited atoms behave as relativistic harmonic oscillators, see \cite{fujiwaraEtal2018NJP}); recently obtained results for the corresponding one-dimensional configuration reproduced the expected trajectories and the period-energy relation with great degree of success \cite{fujiwaraEtal2018NJP}. The extension of such experiments to two-dimensional, anisotropic lattices must exhibit chaotic motion of the excited atoms, according to the results presented here.

\section*{Acknowledgements}
We thank the anonymous referees for the many helpful suggestions which allowed us to improve our manuscript. RSSV thanks Eric Perim for helpful discussions. This work was supported by the S\~ao Paulo Research Foundation (FAPESP), the Brazilian National Research Council (CNPq), and the Coordena\c{c}\~ao de Aperfei\c{c}oamento de Pessoal de N\'ivel Superior - Brasil (CAPES) - Finance Code 001. RSSV acknowledges FAPESP grant 2015/10577-9 and CAPES. This work has made use of the facilities of the Laboratory of Astroinformatics (IAG/USP, NAT/Unicsul), whose purchase was made possible by FAPESP (grant 2009/54006-4) and the INCT-A.




\end{document}